\documentclass[11pt]{JHEP3}

  \usepackage{epsfig,multicol}

  \newcommand{\db}{\bar{d}}

  \newcommand{\bra}{\langle}
  \newcommand{\ket}{\rangle}

  \title{Models for Light-Cone Meson Distribution Amplitudes}

  \author{Patricia Ball and Angelique N. Talbot\\
  IPPP, Department of Physics, 
  University of Durham, Durham DH1 3LE, UK\\
  E-mail: \email{Patricia.Ball@durham.ac.uk},
  \email{A.N.Talbot@durham.ac.uk}}
  \received{}
  \revised{}
  \accepted{}
  \preprint{IPPP/04/83 \\DCPT/04/166}
  \abstract{Leading-twist distribution amplitudes (DAs) 
  of light mesons like $\pi,\rho$
    etc.\ describe
  the leading nonperturbative hadronic contributions to  exclusive QCD
  reactions at large
  energy transfer, for instance electromagnetic form factors.
  They also enter B decay amplitudes described in QCD factorisation, 
  in particular nonleptonic two-body decays. Being
  nonperturbative quantities, DAs cannot be calculated from first
  principles, but have to be described by models. Most
  models for DAs rely on a fixed order conformal
  expansion, which is strictly valid for large factorisation scales, but
  not always sufficient in phenomenological applications. We derive
  models for DAs that are valid to all orders in the conformal expansion
  and characterised by a small number of
  parameters which are related to experimental observables.}


\begin{document}

  \section{Introduction}\label{sec1}

Light-cone distribution amplitudes (DAs) appear in the description of hard
exclusive QCD processes by factorisation, schematically
\begin{equation}\label{eq:1}
\mbox{amplitude~}\sim \prod_{j,i} \phi_{{\rm out},j}(n_j) \otimes
T(Q^2;n_j,n_i)\otimes \phi_{{\rm in},i}(n_i),
\end{equation}
where the labels $i$ and $j$ refer to the hadrons in the incoming and
outgoing states, respectively, $\phi(n)$ is the DA that describes the
probability amplitude for a hadron to be found in the parton state $n$
and $T(Q^2;n_i,n_j)$ is the perturbative function that describes the hard
scattering between the partons; $Q^2$ is the hard momentum transfer. 
The symbol $\otimes$ indicates a
convolution, i.e.\ a sum or integral over the parton degrees of
freedom that correspond to the states $n_i$ and $n_j$. 
Factorisation implies that the process can
be split into two separate regimes, the hard-scattering subprocess
$T$, which involves a short-distance momentum transfer and 
can be calculated in perturbation theory, and the
long-distance hadronisation of the partons emerging from the
short-distance process, described by the process-independent DAs
$\phi$. The expansion of the amplitude in (\ref{eq:1}) is ordered by
inverse powers of the momentum transfer $Q^2$, 
which corresponds to a light-cone
expansion of the process in terms of contributions of
increasing twist. In this paper we are concerned with the
leading-twist DA of light mesons. Although we shall focus on the
pseudoscalar mesons $\pi$ and $K$, the models we suggest are equally
well applicable to --- and in fact have already  \cite{BZ2} been applied to
--- vector mesons ($\rho$, $\omega$, $K^*$ and $\phi$).

The leading-twist meson DA of a meson is related to its
Bethe-Salpeter wave function by integrating out the
dependence on the transversal momentum $k_\perp$,
$$
\phi(u) \sim \int_{k_\perp^2<\mu^2} d^2k_\perp \phi(u,k_\perp),
$$
where $u$ is the longitudinal momentum fraction carried by the quark (and $\bar
u\equiv 1-u$ that carried by the antiquark). Originally introduced in
the context of
hadron electromagnetic form factors or the pseudoscalar-photon
transition form factor \cite{pQCD,BLreport}, 
light-meson DAs have, in recent years,
attracted increasing interest also in B physics 
due to their appearance in QCD sum rules
on the light-cone \cite{BZ2,LCSRs:reviews,protz,BZ} and factorisation
formulas for B decay amplitudes \cite{QCDfac} and form factors
\cite{SCET}. The theory of meson DAs is 
well understood \cite{CZreport,wavefunctionsPS,wavefunctionsV} and
suggests their parameterisation in terms of a partial
wave expansion in conformal spin. One advantage of this expansion is
that the contributions of higher conformal spins to the convolution
integral and hence the physical amplitude are suppressed by the highly
oscillating behaviour of the corresponding partial waves, another one that
conformal symmetry ensures, to leading-logarithmic accuracy, 
multiplicative renormalisation of each
partial wave. The combination of both features suggests 
the construction of models for DAs based on
a truncated conformal expansion, where only the first few waves are
included, typically one to three. Not much is known, in general, about
the amplitudes of these partial waves. For the $\pi$ and $\eta$ 
some experimental
information on the $\pi(\eta)\gamma\gamma^*$ transition form factor is
available \cite{CLEO}, which one can use to extract values of the
first two amplitudes. 
From the theoretical side, there exist a few dated lattice
calculations for the second moment of the $\pi$ DA \cite{lattDA}, and
a recent retry \cite{sachrajda}; unfortunately, these results are
still preliminary and cannot yet be used in phenomenological
applications. Other theoretical calculations, for both pseudoscalar
\cite{CZreport,wavefunctionsPS,Filyanov,SU(3)breaking} 
and vector mesons \cite{wavefunctionsV,BB96}, use the method of QCD
sum rules, which turns out to be not very suitable for higher moments;
we will come back to that point in Sec.~\ref{sec3}.\footnote{These problems
have been evaded in Refs.~\cite{nonlocal,Stefanis2} in a modifed
version of QCD sum rules using nonlocal condensates.} It is probably
fair to say that at present some information is available on the
lowest moments of the $\pi$ DA, but much less so for other mesons. 
In view of the
growing demands on B physics to deliver ``precision results'', cf.\  
e.g.\  Ref.~\cite{bigi}, 
and the prominent role of QCD factorisation
in  testing the mechanism(s) of flavour
violation, it then appears timely to assess the actual theoretical
uncertainty of hadronic decay amplitudes induced by the truncation of
the conformal expansion and to devise alternative models for
light-meson DAs that do not rely on conformal expansion, but
establish a closer connection between the characteristics of DAs
accessible in ``classical'' applications of pQCD in hard exclusive
processes and their use in B
physics. This is precisely what we aim to achieve in this paper. 

Our models for the leading-twist DAs of  the $\pi$
and $K$ are based on the fall-off behaviour of the $n$th Gegenbauer
moment of these amplititudes, $a_n$, 
in $n$, which we assume to be power-like. We shall argue
that such a behaviour can be justified for instance from the known
perturbative contributions to the $a_n$. We formulate our
models in terms of a few parameters, notably the first inverse moment
of the DA, which, for the $\pi$, is directly related to experimental
data, and the strength of the fall-off of the $a_n$. The models
can be summed to all orders in the conformal expansion and give
predictions for the full DA at a certain scale. Depending on the
parameters, the models also
predict a nonstandard behaviour of the DAs at the endpoints, which
affects for instance the scaling, in $m_b$,
of the B$\to\,$light meson form factors
at zero momentum transfer.

Our paper does {\em not} aim to give a fully-fledged analysis
incorporating  all
available constraints on the $\pi$ DA from low-energy experimental
data as scrutinised  in Refs.~\cite{Stefanis2,Schmedding,Stefanis1}. Rather,
we aim to work out the gross features of leading-twist DAs, which are
likely to apply to all light pseudoscalar and vector mesons. 

Our paper is organised as follows: in Sec.~\ref{sec2} we 
define the leading-twist DA of pseudoscalar mesons and discuss its partial-wave
expansion in conformal spin. In Sec.~\ref{sec3} we derive models for
the DA based on the fall-off behaviour of higher partial-waves and
formulate constraints on the model parameters. In
Sec.~\ref{sec4} we investigate the dependence of one important
quantity in B physics, the semileptonic form factor $f_+^{B\to\pi}(0)$,
on the model DAs, and in Sec.~\ref{sec5} we study their effect on CP
asymmetries and branching ratios in nonleptonic B decays. We summarise
and conclude in Sec.~\ref{sec6}.

 \section{Definitions and Conformal Expansion}\label{sec2}

 Distribution amplitudes are defined in terms of matrix elements of 
 nonlocal operators
 near the light-cone. For pseudoscalar mesons $P$ in particular one has
 \begin{eqnarray}
 \langle 0 | \bar q_1(x)\gamma_\mu\gamma_5 [x,-x]
 q_2(-x)|P\rangle & = & i f_P p_\mu \int_0^1 du \, e^{i\xi px} 
 \left[ \phi_P(u) +
   \frac{1}{4}\, m_P^2 x^2 {\mathbb A}_P(u) + O(x^4)\right]\nonumber\\
 &&{} + \frac{i}{2}\,
   f_P m_P^2\,
   \frac{1}{px}\, x_\mu \int_0^1 du \, e^{i\xi px}\,
 \left[{\mathbb B}_P(u)+O(x^2)\right],\hspace*{0.7cm}
 \label{eq:T2}
 \end{eqnarray}
 where $\xi = 2u-1$ and we use
 $[x,y]$ to denote  the Wilson-line
 connecting the points $x$ and $y$:
 \begin{equation}
 [x,y] ={\rm P}\exp\left[ig\!\!\int_0^1\!\! dt\,(x-y)_\mu
   A^\mu(tx+(1-t)y)\right].
 \label{Pexp}
 \end{equation}
 In Eq.~(\ref{eq:T2}), 
 $\phi_P$ is the leading twist-2 DA, whereas $\mathbb A_P$ and $\mathbb B_P$
 contain contributions from higher-twist operators. The corresponding
 definitions of vector meson DAs can be found in Ref.~\cite{wavefunctionsPS}.

  The extraction of the leading behaviour of the matrix elements on
   the light-cone yields ultraviolet divergences, whose regularisation
   generates a nontrivial scale-dependence that can be described by
   renormalisation group methods \cite{pQCD,BLreport}.  
   Conformal invariance of QCD allows one to express the
   DA in terms of a partial wave expansion, also called conformal expansion,
  in contributions from different
   conformal spin, which do not mix with each other under a change of
   scale. This is true to leading
   logarithmic accuracy, but no longer the case at higher order, as the
   underlying symmetry is anomalous. For the leading-twist DA
   $\phi(u)$, for both pseudoscalars and vectors,
  the conformal expansion is in terms of Gegenbauer polynomials
   $C_n^{3/2}$,
   \begin{equation}\label{eq:confexp}
   \phi(u,\mu^2)=6 u (1-u) 
    \sum\limits_{n=0}^\infty a_{n}(\mu^2)  C^{3/2}_{n}(2u-1).
 \end{equation}
 The coefficients $a_n$, the so-called Gegenbauer moments, renormalise
     multiplicatively to  leading
   logarithmic accuracy:
 \begin{equation}
     a_n(Q^2) = a_n(\mu^2)
   \left(\frac{\alpha_s(Q^2)}{\alpha_s(\mu^2)}\right)^{
 \gamma_0^{(n)}/(2\beta_0)}  \label{wf1}
    \end{equation}
    with $\beta_0=11 - (2/3) n_f$. The one-loop anomalous dimension
 is given by \cite{GW}
   \begin{eqnarray}
   \gamma_0^{(n)} &=& 8 C_F
   \left(\psi(n+2)+\gamma_E -\frac{3}{4} - \frac{1}{(n+1)(n+2)}\right).
   \label{eq:1loopandim}
   \end{eqnarray}
 For $\pi$, $\rho$, $\omega$, $\eta$, $\eta'$ and $\phi$, G-parity
 ensures that $a_{\rm odd} = 0$ and that the DA is symmetric under
 $u\leftrightarrow 1-u$, whereas for $K$ and $K^*$ the nonzero values
 of $a_{\rm odd}$ induce an antisymmetric component of the
 DA. $a_0\equiv 1$ is fixed by normalisation,
 $$\int_0^1 \phi(u,\mu^2) = 1,$$
 whereas all other $a_n$ are intrinsically nonperturbative
 quantities. As they do not mix under renormalisation,
 Eq.~(\ref{eq:confexp}) is well suited to construct models for $\phi$:
 truncating the series after the first few terms, typically three,
 yields a parameterisation of the DA that is ``stable'' under a change of
 scale, except for the numerical values of $a_n$. Despite there being
 no small expansion parameter in the game, such a {\em 
 truncated conformal expansion} is often -- but not always -- a
 meaningful approximation to the full DA, as we shall see below.

 As the anomalous dimensions are positive (except for
 $\gamma_0^{(0)}=0$), the contributions of higher $a_n$ get suppressed
 for large scales, and for $\mu^2\to\infty$ the DA approaches the so-called
 asymptotic DA 
 \begin{equation}\label{eq:as}
 \phi_{\rm as}(u) = 6 u (1-u).
 \end{equation}

 For many processes involving DAs, in particular in B physics, it is
 usually argued that a truncated conformal expansion be sufficient for the
 calculation of physical amplitudes as long as the perturbative
 scattering amplitude is ``smooth'' --- the reason being the highly
 oscillatory behaviour of higher order Gegenbauer polynomials.
 It is actually instructive to  quantify this statement, for example for 
 the simplest, but
 phenomenologically very relevant case of one meson in the initial or
 final state, so that the convolution integral reads
 \begin{equation}
 I = \int_0^1 du\,\phi(u) T(u),
 \end{equation}
 where $T$ is the perturbative scattering amplitude.

We distinguish the following cases:
 \begin{itemize}
 \item[(i)] $T$ is nonsingular for $u\in[0,1]$;
 \item[(ii)] $T$ has an integrable singularity at one of the endpoints;
 \item[(iii)] $T$ contains a nonintegrable singularity at one of the
   endpoints.
 \end{itemize}
 As a typical example for case (i), consider $T(u) = \sqrt{u}$, which yields
 $$\int_0^1 du\,\phi(u) \sqrt{u} = \sum_{n=0}^\infty \frac{(-1)^n
   36(n+1)(n+2)}{(2n-1)(2n+1)(2n+3)(2n+5)(2n+7)} \, a_n.$$
 This result implies a strong fall-off $\sim 1/n^3$ of the coefficients of
   higher Gegenbauer moments $a_n$: assuming $a_i\equiv 1$ for all $i$, already
   the first three terms in the sum account for 98.8\% of the full
   amplitude. In reality the convergence will be even better as all
   existing evidence points at $|a_n|\ll 1$ for $n\geq 1$. 

 As an example for case (ii) choose $T(u) = \ln u$, which yields 
 $$\int_0^1 du\,\phi(u) \ln(u) = -\frac{5}{6}\, a_0 + \sum\limits_{n=1}^\infty
 \frac{(-1)^{n-1}}{n(n+3)}\, 3 a_n.$$
 The singularity at $u=0$ evidently worsens the convergence of the
 series; again assuming $a_i\equiv 1$, the first three terms now
 overshoot the true result by 35\%. In order to approximate the full
 amplitude to within 5\% one now has to include nine terms, but the
 convergence will again be better in practice, thanks to the fall-off
 of $a_n$ in $n$.

 Case (iii) is more complicated and  depends on the
 asymptotic behaviour of the $a_n$. For $T(u)=1/u$, for instance, one obtains
 \begin{equation}\label{nn}
 \int_0^1 du\,\phi(u)\,\frac{1}{u} = 3 \sum_{n=0}^\infty (-1)^n a_n.
 \end{equation}
Here the amplitude is finite only if the $a_n$ fall off sufficiently
 fast in $n$.
 For stronger endpoint divergences the coefficients multiplying $a_n$
 start to grow in $n$ and for $T\sim 1/u^2$ the integral diverges, even
 for the asymptotic DA, which would indicate a breakdown of
 factorisation for that process.

 The conclusion to be drawn from this discussion is that models for
 $\phi$, based on a conformal expansion that is truncated after the
 first few terms, are indeed appropriate for cases (i) and (ii), but
 less so for case (iii). Convolutions with $T\sim 1/u$ are actually
 very relevant both in hard perturbative QCD reactions, e.g.\
 $\gamma\gamma^*\to\pi$, and in B physics, e.g.\ $B\to\pi\pi$
 \cite{QCDfac}. Given the different weight the $a_n$ do have in
 different convoluted amplitudes, and the fact that their impact is
 highest in convolutions of type (iii), it appears not unreasonable to
 base a parameterisation of $\phi$ not on, say, $a_2$ and $a_4$,
 setting all $a_{n>4}=0$ , but
 rather on the ``worst case scenario'' of Eq.~(\ref{nn}), where all
 $a_n$ enter with the highest possible weight factor. This is the
 basic idea of our models of leading-twist DAs we shall elaborate on
 in the next section.

 \section{\boldmath Models for $\phi_\pi$ and $\phi_K$}\label{sec3}

 There are basically two properties that any viable model for the 
 leading-twist $\pi$ DA $\phi_\pi$ must fulfill:
 \begin{itemize}
 \item[(a)] in the limit of large energies, $\mu^2\to\infty$, the DA must
   approach the asymptotic DA $\phi_\pi(u,\mu^2=\infty) = 
6 u \bar u$ \cite{pQCD};
 \item[(b)] the first inverse moment, $\int_0^1 du\,\phi_\pi(u)/u$, which
   is related to the $\pi\gamma\gamma^*$ transition form factor,
   must exist.
 \end{itemize}
Condition (a) must evidently be fulfilled also for all other
 light-meson DAs; condition (b) should be fulfilled in general 
if QCD factorisation in B decays \cite{QCDfac} 
is to make sense. Obviously any model that is based on a truncated
 conformal expansions fulfills both constraints, and in
 addition predicts that for $u\to 0,1$, $\phi\sim u(1-u)$ independent of
 the factorisation scale. We shall show that this prediction is, in
 general, not fulfilled for models that are not truncated at fixed order
 in the conformal expansion.

 The starting point for our models is the relation between the first
 inverse moment of $\phi$ and the sum of all Gegenbauer moments:
 \begin{equation}
 \int_0^1 du\,\frac{\phi(u,\mu^2)}{3u} 
\equiv \Delta(\mu^2) = 1 + \sum_{n=1}^\infty
 (-1)^n a_{n}(\mu^2).
 \end{equation}
 As mentioned before, one has to distinguish between mesons for which
 $a_{\rm odd}$ vanish due to G-parity, such as the $\pi$, and strange
 mesons, for which the odd moments induce an antisymmetric part of the
 DA. It is the former case we shall study first.
 Available experimental data for the $\pi$, as summarised in
 Ref.~\cite{Stefanis1},  
 point at a value of $\Delta$ around $1.1$
 at the scale $\mu\approx1.2\,$GeV, which implies that the infinite sum
 be convergent. Hence, even at the comparatively low scale 1.2$\,$GeV the
 $a_{2n}$ must fall off fast enough in $n$. Assuming an asymptotic
 equal-sign behaviour for large $n$, a power-like fall-off as
 $a_{2n}\sim 1/n^p$ with $p$ slightly larger than 1 is one of the
 slowest possible fall-offs. It turns out that
 DAs defined by a power-like fall-off of
 the Gegenbauer moments can actually be summed explicitly: using the
 generating function of Gegenbauer polynomials,
 $$
 f(\xi,t) = \frac{1}{(1-2 \xi t + t^2)^{3/2}} = \sum_{n=0}^\infty
 C_n^{3/2}(\xi)\, t^n,
 $$
 the DA with moments
 $$a_{n} = \frac{1}{(n/b+1)^a}, \quad \mbox{for $n$ even},$$
 is given by
 $$\tilde\phi_{a,b}^+(u) = \frac{3u \bar u }{\Gamma(a)}\,
 \int_0^1 dt (-\ln t)^{a-1}\, \left( f(2u-1,t^{1/b})
 + f(2u-1,-t^{1/b})\right).
 $$
 In the same way one obtains a DA with alternating-sign behaviour of
 the Gegenbauer moments,
 $$a_{n} = \frac{(-1)^{n/2}}{(n/b+1)^a}, \quad \mbox{for $n$ even},$$
 as
 $$\tilde\phi_{a,b}^-(u) = \frac{3u \bar u }{\Gamma(a)}\,
 \int_0^1 dt (-\ln t)^{a-1}\, \left( f(2u-1,it^{1/b})
 + f(2u-1,-it^{1/b})\right).
 $$
 The corresponding values of $\Delta$ are $\Delta_{a,b}^+ =
 (b/2)^a \zeta(a,b/2)$ and $\Delta_{a,b}^- = (b/4)^a (\zeta(a,b/4) -
 \zeta(a,1/2+b/4))$, 
 where $\zeta(a,s)= \sum_{k=0}^\infty 1/(k+s)^a$ is the Hurwitz $\zeta$
 function.
 In order to obtain models for arbitrary
 values of $\Delta$, we split off the asymptotic DA and write
 \begin{eqnarray}\label{model}
 \phi^\pm_{a,b}(\Delta) &=& 6 u \bar u + \frac{\Delta-1}{\Delta_{a,b}^\pm -1}
 \left(\tilde{\phi}^\pm_{a,b}(u) - 6 u \bar u\right), \quad\mbox{valid
   for $a\geq 1$ and $b>0$}.
 \end{eqnarray}
Is there any reason why the $a_n$ should fall off as inverse powers in
$n$ --- other than that the corresponding DAs can be expressed in
closed form via Eq.~(\ref{model})? The answer to this question relies
on the behaviour of $a_n(\mu^2)$ under a change of scale.
Evidently the models are defined at one particular scale, for instance
 the hadronic scale $\mu\approx 1.2\,$GeV. At larger scales $Q^2$, the
 $a_n$ change according to Eq.~(\ref{wf1}). For large $n$, we have
 $$\gamma_0^{(n)}\stackrel{n\to\infty}{\approx} 8 C_F \ln n + O(1)
 $$
 and 
 $$ a_n(Q^2) \stackrel{n\to\infty}{\approx} \frac{1}{n^{4 C_F/\beta_0\,
     \ln (1/L)}} \,a_n(\mu^2)$$
 with $L = \alpha_s(Q^2)/\alpha_s(\mu^2)$. That is: perturbative
     leading-order scaling induces a power-like fall-off of the $a_n$,
     at least for large $n$. We take this as indication that such a
     behaviour is indeed intrinsic to QCD. One more consequence of
     scaling is that for $Q^2\to\infty$,
     i.e.\ $L\to 0$, the suppression of higher $a_n$ is power-like
     with a power that approaches infinity, so that
 $$\tilde{\phi}^{\pm}(Q^2\to\infty) = \tilde{\phi}^\pm(a\to\infty) = 6 u
     (1-u).$$
Hence $\phi_{a,b}^\pm$ as defined in
     (\ref{model}) approaches the asymptotic DA in this limit,
which implies that condition (a) follows from (b). 

Eq.~(\ref{model}) implies that the asymptotic DA is recovered 
for $\Delta=1$, and also from
 $\phi_{a,b}^+$ in the limit $a\to 1$. The models are valid only for
 $a\geq 1$, as otherwise $\Delta^+_{a,b}$ diverges, or, equivalently,
 $\phi_{a,b}^\pm$ does not vanish at the endpoints $u=0,1$. 
 In
 Fig.~\ref{fig:2} we plot several examples of $\phi_{a,b}^\pm$ for a fixed
 value of $\Delta$; it is evident that the two models
 $\phi_{a,b}^+$ and $\phi_{a,b}^-$ have a rather dissimilar functional
 dependence on $u$; in particular $\phi_{a,b}^-$ turns out to be
 nonanalytic at $u=1/2$ for $a\leq 3$.  The
 dependence of $\phi_{a,b}^\pm$ on $b$ is illustrated in
 Fig.~\ref{fig:b}.  In Fig.~\ref{fig:3} we show the
 possible values of the lowest Gegenbauer moments $a_{2,4}$ that can be
 obtained for fixed values of $\Delta$ and $b$, but different $a$. For
 $\phi_{a,b}^+$, one always has $a_2<\Delta-1$, for $\phi_{a,b}^-$,
 $a_2>\Delta-1$; in the limit $a\to\infty$ both branches meet and one
 obtains the truncated NLO conformal expansion with $a_2=\Delta-1$
 and $a_4=0$.

 \FIGURE{
 $$\epsfysize=0.3\textwidth\epsffile{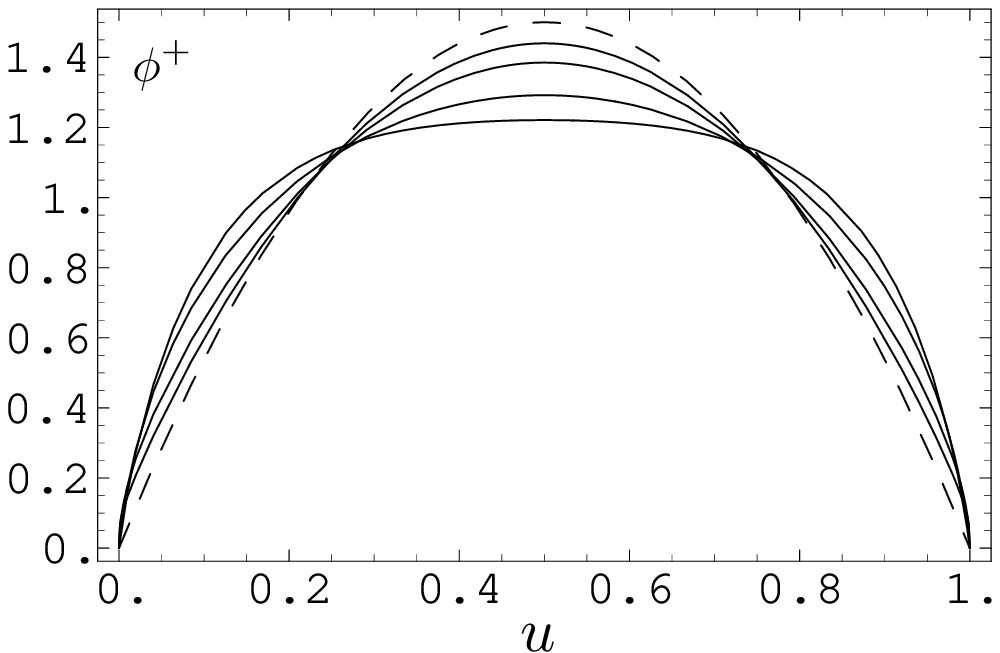}\qquad
 \epsfysize=0.3\textwidth\epsffile{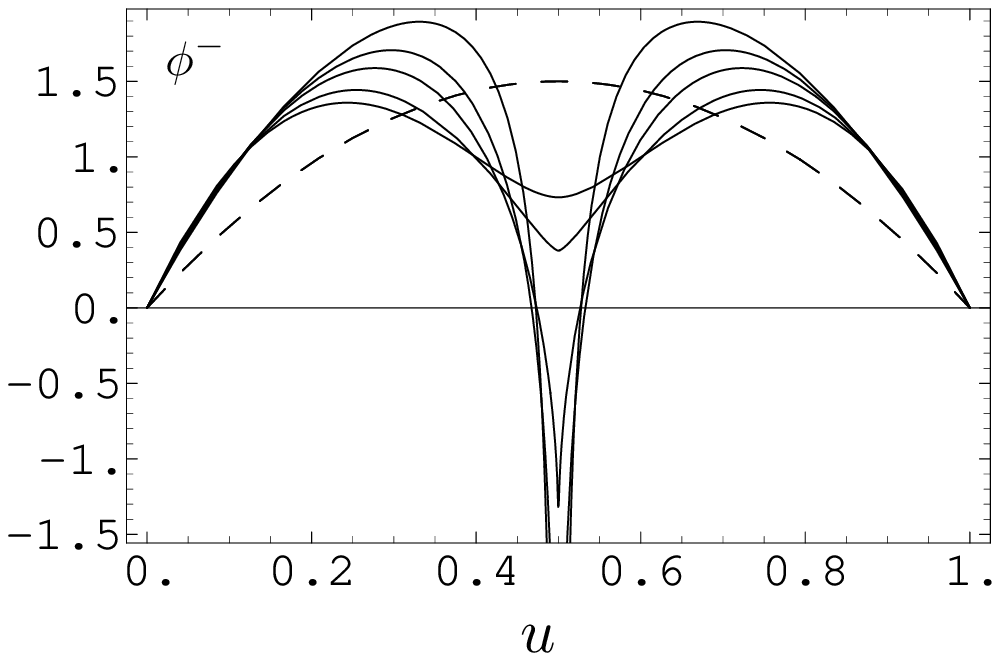}$$
 \caption[]{Left: Examples for model DAs $\phi_{a,b}^+$ as 
functions of $u$, for
   $a=1.5,2,3,4$, $b=2$ and $\Delta=1.2$ 
 (solid curves), as compared to the asymptotic DA
   (dashed curve). For $a\to 1$, $\phi^+_{a,b}$  approaches the asymptotic
   DA. Right: the same for $\phi_{a,2}^-$. Note that for $a\leq 3$,
   $\phi_{a,b}^-$ is nonanalytic at $u=1/2$ and displays a pronounced
   ``spike''.}\label{fig:2}}

 \FIGURE{
 $$\epsfysize=0.3\textwidth\epsffile{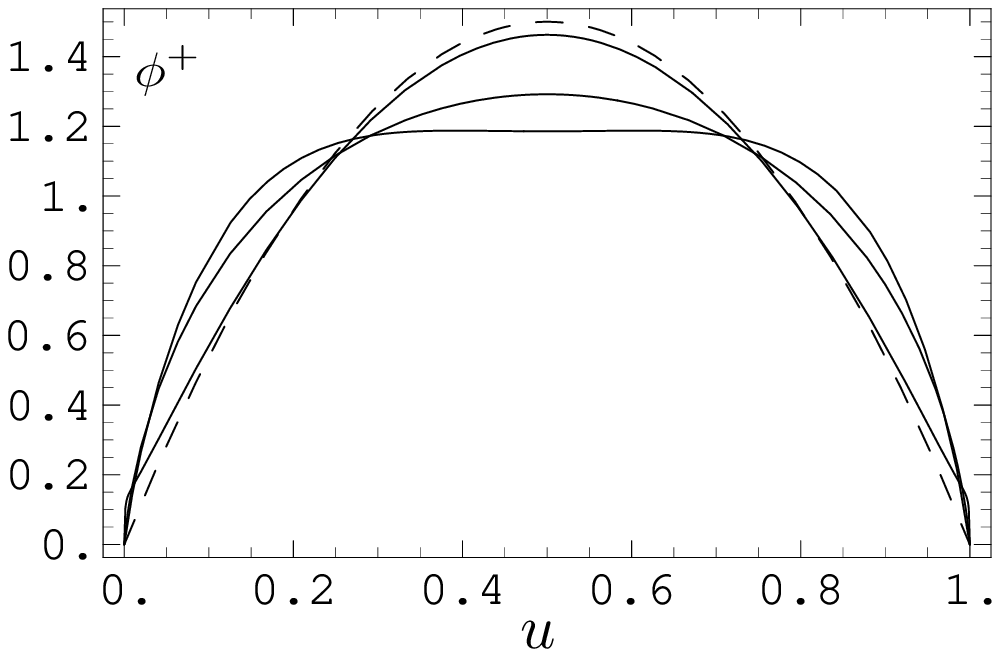}\qquad
 \epsfysize=0.3\textwidth\epsffile{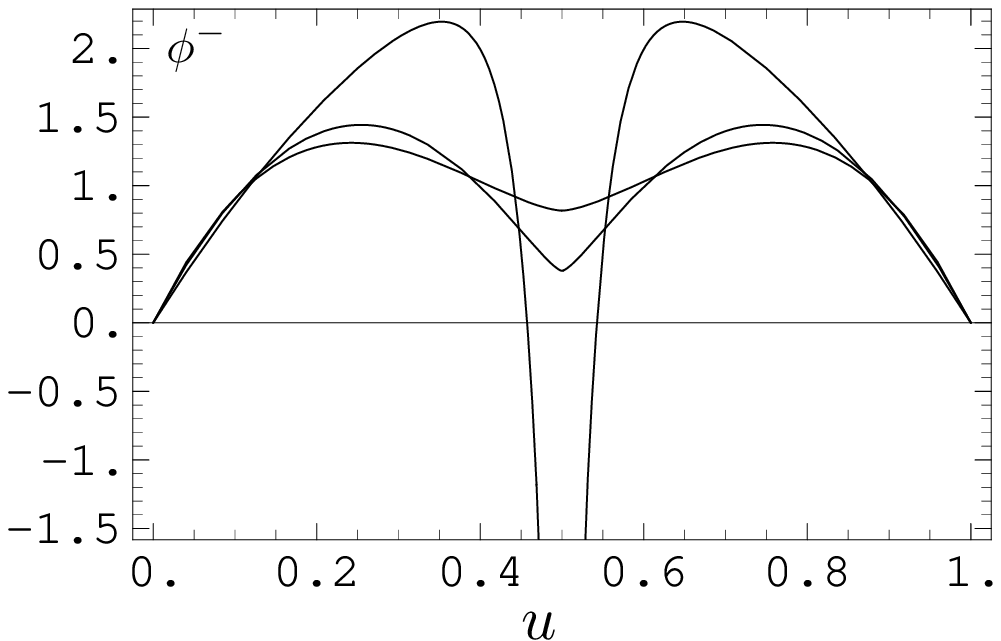}$$
 \caption[]{Left: Examples for model DAs $\phi_{3,b}^+$ as functions
   of $u$, for
   $b=0.2,2,20$ and $\Delta=1.2$ 
 (solid curves), as compared to the asymptotic DA
   (dashed curve). For $b\to \infty$, $\phi^+_{a,b}$  approaches the asymptotic
   DA, as $\Delta^+_{a,\infty}\to\infty$. 
 Right: the same for $\phi_{3,b}^-$.}\label{fig:b}
 }

 \DOUBLEFIGURE{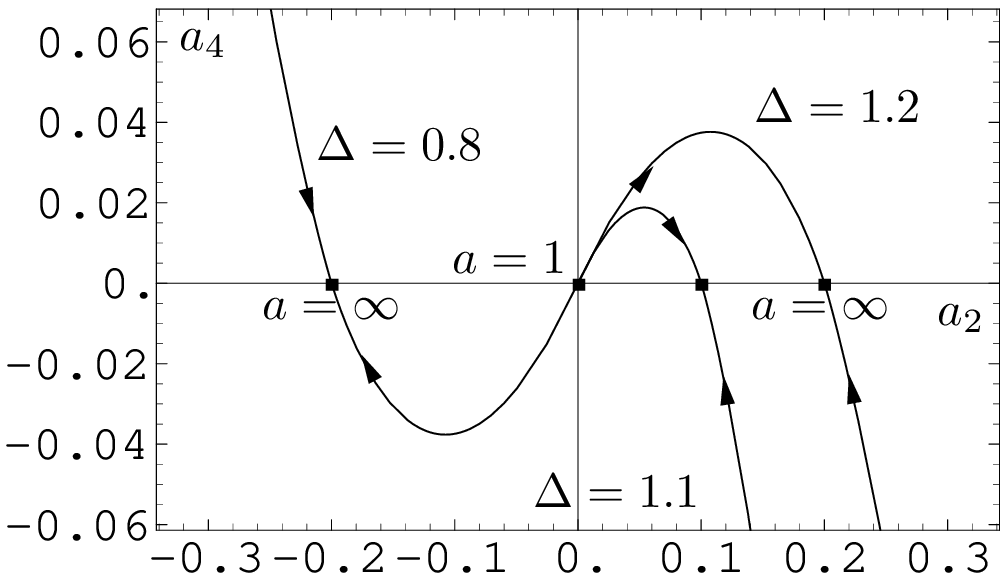,height=0.27\textwidth}{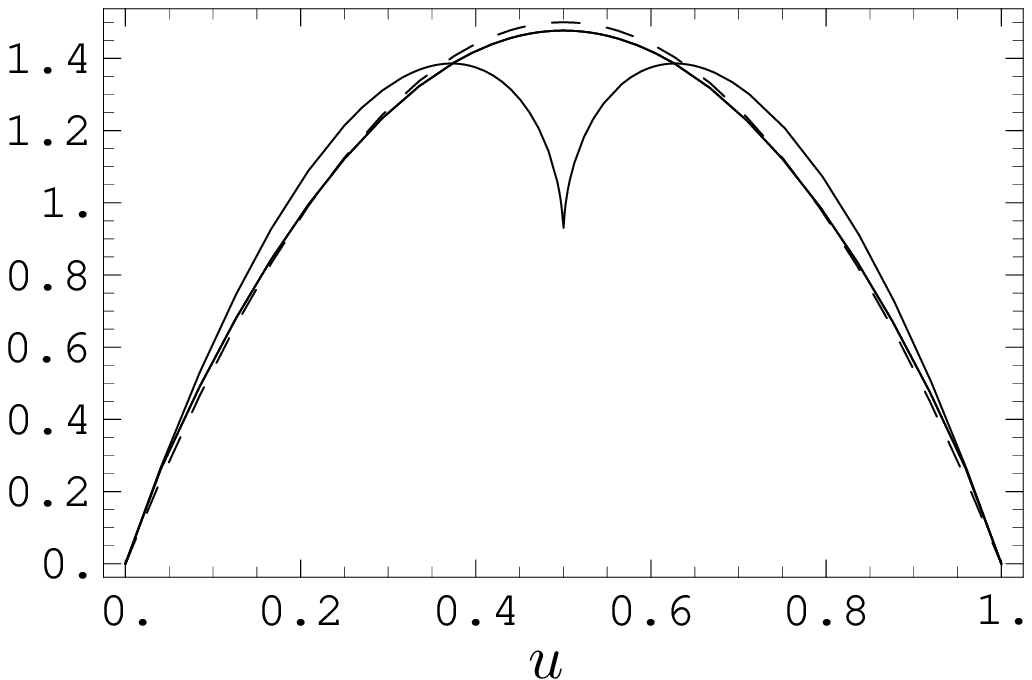,height=0.3\textwidth}{Values of $a_2$ and $a_4$ for different parameters $\Delta$
 and $a$ of $\phi^\pm_{a,2}$. 
 The curves in the upper right and lower left quadrant
 correspond to $\phi^+_{a,2}$, those in the lower right and upper left
 one to $\phi^-_{a,2}$.\label{fig:3}}{Comparison of $\phi^\pm_{2,2}(u)$ 
 (solid curves) with the
   asymptotic DA (dashed curve) for $\Delta=1.04$. The ``spike'' at
   $u=1/2$ characteristic for $\phi^-$ causes these models to
   significantly deviate from the asymptotic DA even for $\Delta$
   close to 1.\label{fig:deviate}}

 Before discussing constraints on the model parameters $\Delta$, $a$
 and $b$, let us shortly comment on the behaviour of the DAs near the
 endpoints $u=0,1$. In many applications of DAs it is assumed --
 implicitly or explicitly -- that $\phi\sim u (1-u)$ near the
 endpoints. Although this is evidently the case for any model based on
 a truncated conformal expansion, it does not apply to our models, at
 least not for all values of $a$. A closer inspection shows that
 $\phi^+\sim \sqrt{u(1-u)}$ for $a=2$ and $\sim u (1-u) \ln (u(1-u))$
 for $a=3$. It is only for $a>3$ that the DAs behave in the
 ``canonical'' linear way. Are there are 
any rigorous arguments why the DAs
 should behave linearly near the endpoints, even for low hadronic
 scales? As far as we could trace the origin of the argument in favour
 of linear behaviour, it was stated first
in Ref.~\cite{CZreport} and relies on the fact that QCD sum rules for 
the moments of the DA exhibit the following behaviour:
 $$\langle \xi^n\rangle = \int_0^1 \xi^n \phi(u)
 \stackrel{n\to\infty}{\sim} \frac{1}{n^2}.$$
 Two comments are in order here. First, an exact calculation for arbitary $\phi$ shows that
 $$\langle \xi^n\rangle = \frac{1}{4n^2}\,\left(\phi'(0) -
 \phi'(1)\right) + O(n^{-3}),$$
 so that the correct statement is that $\langle \xi^n\rangle\sim
 1/n^2$ implies that $\phi'(u)$ exists at the endpoints -- which in
 turn indeed implies a linear behaviour, since $\phi(0)=0=\phi(1)$ as long
 as $\Delta<\infty$. Second, the conclusion of the authors of
 \cite{CZreport} is based on results obtained from the leading-order 
perturbative contributions to QCD sum rules. NLO expressions have been
 obtained in
 \cite{SU(3)breaking,BB96} 
 and yield
 $\phi\sim u (1-u) \ln^2(u/(1-u))$,  which upsets the linear approach
 to the endpoints and is equivalent to $a_n\sim 1/n^3$. 
The large-$n$ behaviour of nonperturbative terms
 cannot be obtained from these sum rules,\footnote{The reason being
   that $a_n$ for large $n$ are intrinsically nonlocal quantities,
   which cannot be  obtained 
 from a local operator product expansion (OPE). The ineligibility of the
 local OPE manifests itself as contributions to $a_n$ that scale as
 {\em positive} powers in $n$, which is incompatible with a finite
 value of $\Delta$.} but we see no reason why it should not follow the
 $a_n\sim 1/n^3$ behaviour of NLO perturbation theory or even introduce $1/n^2$
 scaling. We hence conclude that the standard assumption of $\phi\sim u
 (1-u)$ is not rigorously justified at hadronic scales
and that all corresponding
 conclusions, in particular the scaling behaviour of the $B\to\,$light
 meson form factors at zero momentum transfer, $f(0)\sim 1/m_b^{3/2}$,
   first derived in \cite{first}, should be taken {\em cum grano
    salis.} We also would like to add one more remark about the
expectation that for small $\Delta-1$ the DA should be ``very
  close'' to the asymptotic DA. In Fig.~\ref{fig:deviate} we show that
  this statement is indeed true for $\phi^+$ models, but not for
  $\phi^-$, which are characterised by their nonanalytic behaviour at
  $u=1/2$. Of course the asymptotic DA is recovered for $\Delta-1\to
 0$, but not in an analytic way.

 \TABLE{
 \begin{tabular}{|l|llllll|}
 \hline
 $a$ & 2 & 3 & 4 & 5 & 6 & $\infty$\\\hline
 $\Delta^+_{\rm max}$ & 2.04 & 1.58 & 1.43 & 1.36 & 1.33 & 1.27\\
 $\Delta^-_{\rm max}$ & 1.04 & 1.11 & 1.16 & 1.19 & 1.22 & 1.27\\\hline
 \end{tabular}
 \caption[]{Upper bound on $\Delta^\pm$, for various values of $a$, as
   implied by $\phi_\pi(1/2)>0.9$. The reference scale is $\mu\approx
   1.2\,$GeV.}\label{tab1}
 }

 Let us now discuss the experimental restrictions on $\Delta$, $a$ and $b$ for
 the $\pi$. First, we fix the reference scale as $\mu = 1.2\,$GeV.  As
 the data are too scarce to constrain all parameters in a meaningful
 way, we fix $b\equiv 2$. As for $\Delta$, we require it to be
 larger or equal 1. This is the case if $a_2\geq 0$, which is indeed
 the common overall conclusion of all determinations available in the
 literature, as compiled in Ref.~\cite{Stefanis1}. An upper bound on
 $\Delta$ can be inferred from experimental data or from requiring
 $a_2\leq 0.2$, which is again the maximum value allowed by most 
 analyses.\footnote{With the notable exception of Chernyak and Zhitnitsky
 \cite{CZreport}, whose results are however by now generally considered to be
 excluded by experiment.} For $\phi^+$ the restriction on $a_2$ implies
 $\Delta \leq 1.2$, which coincides with the allowed range of $\Delta$
 found in \cite{Stefanis2}, whereas for $\phi^-$ the resulting range of
 $\Delta$ is smaller. One more constraint on $\phi^\pm$ is the allowed
 range of $\phi_\pi(1/2)$, which follows from light-cone sum rules for
 the $\pi NN$ coupling  \cite{Filyanov}:
 $$0.9\leq \phi_\pi (1/2,1\,{\rm GeV}) \leq 1.5.$$
 For $\Delta>1$, $\phi(1/2)$ is always smaller than 1.5, so it is only
 the lower bound that is relevant. In Tab.~\ref{tab1} we list the
 corresponding maximum $\Delta$ for various $a$.
 Evidently the constraints on $\Delta^+$ are weaker than the ones
   discussed before, but for $\phi^-$ the minimum value of
   $\phi_\pi(1/2)$ poses a nontrivial constraint on $\Delta$.

 It is possible to further refine the constraints for the $\pi$ 
  as it was done
 in e.g.\ \cite{Stefanis2,Schmedding}. In this paper, however, we are
  not so much interested in the $\pi$, but rather in
 the gross characteristics of the DAs, which are likely to be valid
 also for
 other pseudoscalar and vector mesons.  We hence refrain from pursuing that
 line of investigation, but summarise the main constraints on the
  symmetric parts of  
 our model DAs, which are likely to be valid also for other mesons:
 \begin{itemize}
 \item $1\leq \Delta\leq 1.2$ for $0\leq a_2\leq 0.2$ for
 $\phi^+_{a,2}$: this is based on the observation \cite{CZreport} that
 DAs of mesons with higher mass tend to become narrower; 
 \item $1\leq{\rm Min}(1.2,\Delta^-_{\rm max})$  for $\phi^-_{a,2}$,
 with $\Delta^-_{\rm max}$ given in Tab.~\ref{tab1};
 \item $b=2$, lacking further data.
 \end{itemize}

 Let us now turn to $\phi_K$, the twist-2 DA of the $K$ meson. It
 differs from $\phi_\pi$ by the fact that now also odd Gegenbauer
 moments contribute. 
 Models for the antisymmetric part of the DA can be
 constructed in a similar way as before as\footnote{In complete analogy
   to the symmetric part of the DA we could introduce one
   more parameter $d$ that would correspond to $b$. In view of the near
   complete absence of any information on antisymmetric DAs, we refrain
   from writing down the corresponding formulas and set $d\equiv 2$
   from the very beginning.}
 \begin{eqnarray}
 \tilde\psi_c^+(u) &= &\frac{3u \bar u }{\Gamma(c)}\, 
 \int_0^1 dt (-\ln t)^{c-1}\, \left( f(2u-1,\sqrt{t})
 - f(2u-1,-\sqrt{t})\right),\nonumber\\
 \tilde\psi_c^-(u) &= &\frac{3u \bar u }{i\Gamma(c)}\, 
 \int_0^1 dt (-\ln t)^{c-1}\, \left( f(2u-1,i\sqrt{t})
 - f(2u-1,-i\sqrt{t})\right).
 \end{eqnarray}
 The models are characterised by $c$ and yield $a_1=(2/3)^c$.
 Models for the antisymmetric part of $\phi$ with arbitrary $a_1$ can then
 be defined as
 \begin{equation}
 \psi_c^{\pm} = a_1 (3/2)^c \tilde\psi_c^\pm(u).
 \end{equation}
 Examples for such models are shown in Fig.~\ref{fig:4}.

 \FIGURE{
 $$\epsfysize=0.3\textwidth\epsffile{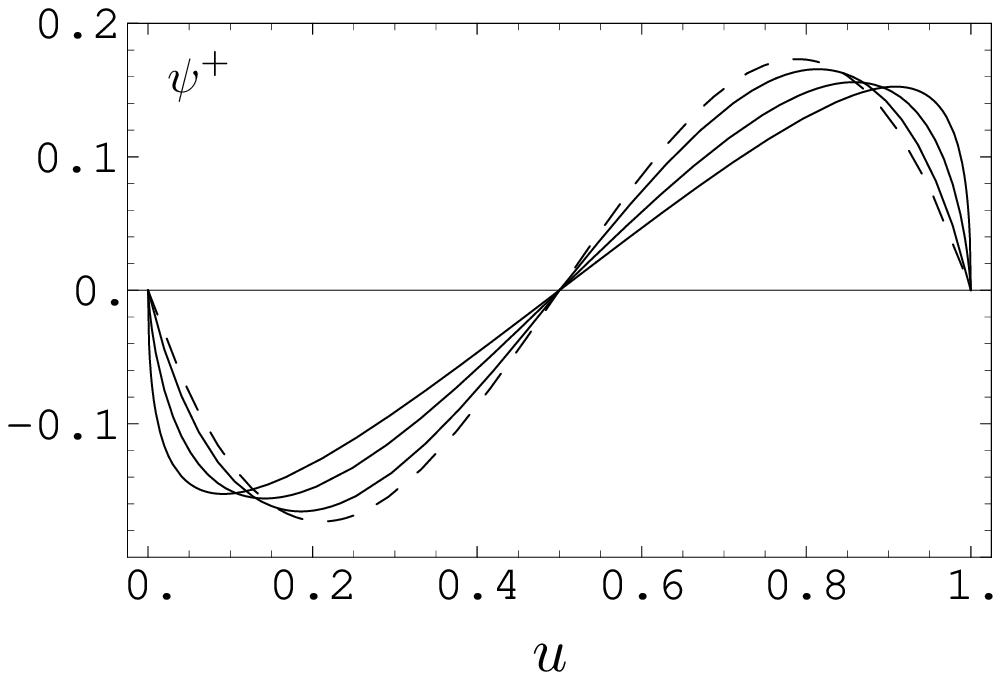}
 \qquad\epsfysize=0.29\textwidth\epsffile{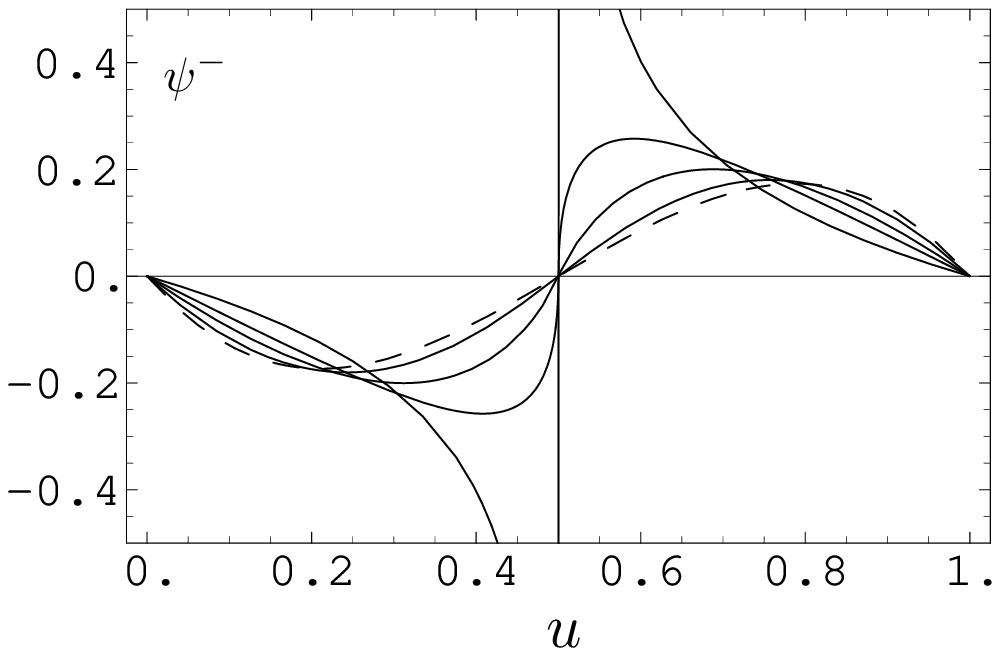} $$
 \caption[]{Models for the antisymmetric contributions to the twist-2
   DA for $a_1 = 0.15$. Left: $\psi_c^+$ as function of $u$ for
   $c\in\{2,3,5\}$ (solid curves), dashed curve: $\psi_\infty^+$. Right: 
 $\psi_c^-$ as function of $u$ for
   $c\in\{1,2,3,5\}$ (solid curves), dashed curve: $\psi_\infty^-$. Like
   the symmetric models $\phi_a^-$, $\psi_c^-$ 
 is nonanalytic at $u=1/2$ for $c\leq 3$. }\label{fig:4}}

 Numerical values for $a_1^K$ have been discussed in \cite{BZ}; we
 quote
 $a_1^K(1.2\,{\rm GeV}) \approx 0.15.$
 For $\phi_K$, the total $\Delta_{\rm tot}$ can then be written as
 $$\Delta^{{\rm tot},\pm} = \Delta + \Delta^{{\rm asym},\pm},$$
where  $\Delta=\sum a_{\rm even}$ is the contribution of the symmetric
part of the DA to $\Delta^{\rm
    tot}$ and  $\Delta^{{\rm asym},\pm}$ is defined as
 \begin{eqnarray*}
\Delta^{{\rm asym},+} & = & \int_0^1 du\,\frac{\psi_c^+(u)}{3u} = - a_1 (3/2)^c
\zeta(c,3/2),\\
 \Delta^{{\rm asym},-} & = & \int_0^1 du\,\frac{\psi_c^-(u)}{3u} = - a_1 (3/4)^c
\left\{\zeta(c,3/4)-\zeta(c,5/4)\right\}.
 \end{eqnarray*}

\section{\boldmath Results for $f_+^{B\to\pi}(0)$}\label{sec4}

One important application of factorisation in B physics is the
calculation of the weak decay form factor $f_+^{B\to\pi}$ from QCD sum
rules on the light-cone. It is beyond the scope of this paper to
review the method of QCD sum rules on the light-cone (LCSRs), for which we
refer to Ref.~\cite{LCSRs:reviews}. Instead, we would like to stress
that LCSRs offer a means to calculate nonperturbative hadronic
quantities like form factors within a controlled approximation,
which relies on the factorisation of an unphysical correlation
function, whose imaginary part is related to the hadronic quantity
in question.

The $B\to\pi$ form factors are defined as 
\begin{equation}
\langle \pi(p)|V_{\mu}|B(p_B) \rangle = \left[(p+p_B)_{\mu}-
\frac{m_B^2-m_\pi^2}{q^2}\,q_{\mu}\right]
f^{B\to\pi}_+(q^2)+\frac{m_B^2-m_\pi^2}{q^2}\,q_{\mu}
\,f^{B\to\pi}_0(q^2),
\end{equation}
with $q=p_B-p$. 
Within the LCSR method $f_{+,0}^{B\to\pi}$ can be related to a
correlation function
\begin{eqnarray}
\label{eq:corr}
\Pi_{\mu}(q,p_B) &=& i\int d^4x e^{i q\cdot x} \bra \pi(p)|
T V_{\mu}(x) j_B^{\dagger}(0) |0 \ket \\
&=& \Pi_+(q^2,p_B^2)(p+p_B)_{\mu} + \Pi_-(q^2,p_B^2) q_{\mu} \quad ,\nonumber
\end{eqnarray}
where $j_B= m_b \db i\gamma_5 b$ is the interpolating field for the
$B$ meson. For unphysical $p_B^2\ll m_b^2$, $\Pi_\pm$ can be expanded
around the light-cone as 
\begin{equation}
\label{eq:lcexp}
\Pi^{\rm LC}_\pm(q^2,p_B^2) = \sum_n \int_0^1 du \,
T_\pm^{(n)}(u,q^2,p_B^2,\mu) \phi^{(n)}(u,\mu),
\end{equation}
where the sum runs over contributions of increasing twist and the term
in $n=2$ is just the leading-twist contribution.
$T_\pm^{(2,3)}$ are known to $O(\alpha_s)$ \cite{BZ}, whereas $T_\pm^{(4)}$ is
known at tree-level \cite{protz}. We have extended the calculation
performed in \cite{BZ} to include contributions up to $a_{10}$. 
The LCSR for the form factor $f_+$ depends on the spectral density
$\rho_+^{\rm LC}$ of $\Pi_+^{\rm LC}$ in $p_B^2$ and reads 
\begin{equation}\label{srx}
e^{-m_B^2/M^2} m_B^2f_B\;f^{B\to\pi}_+(q^2) = 
\int_{m_b^2}^{s_0} ds\, e^{-s/M^2}
\rho_+^{\rm LC}(s,q^2),
\end{equation}
where $M^2$, the so-called Borel parameter, and $s_0$, the
 continuum-threshold, are sum rule specific parameters. Using the
 central values of these parameters as obtained in \cite{BZ},
 $M^2=9.2\,{\rm GeV}^2$ and $s_0=33.9\,{\rm GeV}^2$, and the model
 twist-2 DAs $\phi^\pm_{a,2}$ we obtain the values of
 $f_+^{B\to\pi}(0)$ shown in Fig.~\ref{fig:fpl0}.
\FIGURE{$$\epsfysize=0.31\textwidth\epsffile{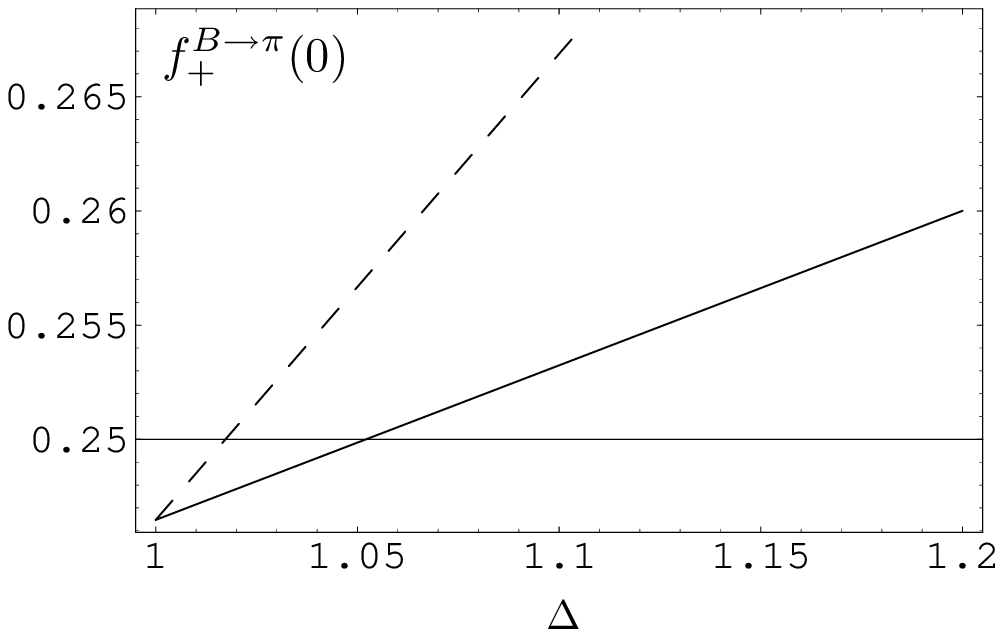}\quad
  \epsfysize=0.31\textwidth\epsffile{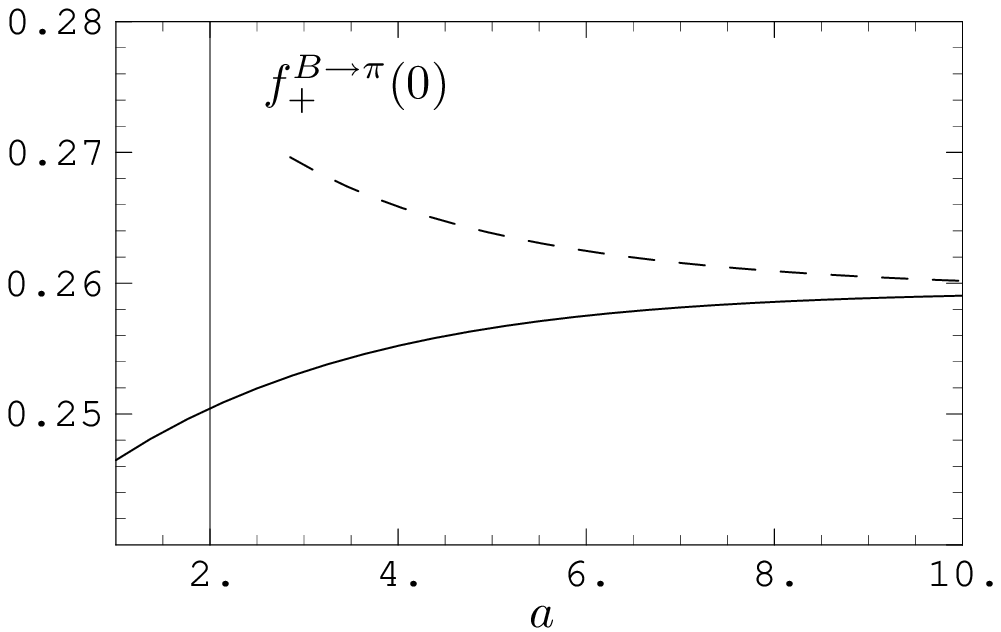}$$
\caption[]{$f_+^{B\to\pi}(0)$ calculated from LCSRs, 
using the twist-2 DAs $\phi_{a,2}^+$ (solid lines) and $\phi_{a,2}^-$ 
(dashed lines). Left: $f_+^{B\to\pi}(0)$ as function of $\Delta$ for
$a=3$ fixed; right: $f_+^{B\to\pi}(0)$  as function of $a$ for 
$\Delta=1.1$ fixed. The endpoints of the dashed curves are set by the
constraint $\phi(1/2)\geq 0.9$.}\label{fig:fpl0}}
The dependence of $f_+^{B\to\pi}(0)$ on $\Delta$, with $a$ fixed, is
linear, as the values of the individual $a_{n\geq 2}$ are just
rescaled by a common factor. If, on the other hand, $\Delta$ is kept
fixed and $a$ is being varied, higher $a_n$ are increasingly
suppressed with increasing $a$, so that for $a\to\infty$ the form
factor obtained using $\phi^+$ approaches that obtained using
$\phi^-$, as is clearly visible in Fig.~\ref{fig:fpl0}. 

Figure~\ref{fig:fpl0} shows that the theoretical uncertainty induced
by $\Delta$ and $a$ is about $\pm 5\%$. This has to be compared with
the final value of $f_+^{B\to\pi}(0)$ stated in Ref.~\cite{BZ}:
$0.258\pm 0.031$, i.e.\ a 12\% uncertainty, which includes also the
variation of other input parameters of the LCSR. 

\section{\boldmath $B\to\pi\pi$ and $B\to
  K\pi$ in QCD Factorisation}\label{sec5}

Another important application of leading-twist DAs are nonleptonic B
decays treated in QCD factorisation \cite{QCDfac}. Recent experimental
data point at a failure of QCD factorisation to explain the observed
branching ratios and CP asymmetries in $B\to\pi\pi$ and $B\to\pi K$,
which has prompted a number of authors to explain this effect by new
physics, e.g.\ Ref.~\cite{newphys}, 
but has also motivated other authors to investigate the
impact of nonfactorisable corrections to QCD factorisation, which are
suppressed by inverse powers of $m_b$ \cite{Feldmann}. In this section
we investigate the effect of nonstandard DAs on the predictions of QCD
factorisation. The factorisation formulas depend on twist-2 DAs of the
$\pi$ and $K$, but also on the form factor $f_+^{B\to\pi}(0)$, for
which we use the results from LCSRs, cf.\ Sec.~\ref{sec4}.

Let us first study the time-dependent 
CP-asymmetry in $B\to\pi^+\pi^-$, which is defined as
\begin{eqnarray}
  A_{CP}^{\pi\pi}&=&\frac{\Gamma(\bar B_0\to\pi^+\pi^-)-
\Gamma(B_0\to\pi^+\pi^-)}
  {\Gamma(\bar B_0\to\pi^+\pi^-)+\Gamma(B_0\to\pi^+\pi^-)} =
  S_{\pi\pi} \sin \Delta m t + C_{\pi\pi} \cos \Delta m t,
\end{eqnarray}
where our interest is in the mixing-induced asymmetry $S_{\pi\pi}$ which
depends on the unitarity triangle (UT) angles $\beta$ and $\gamma$ via
\begin{equation}
  S_{\pi\pi} =\frac{2\textrm{Im}\lambda_{\pi\pi}}{1+|\lambda_{\pi\pi}|^2},
   \qquad 
\lambda_{\pi\pi}=e^{-2i\beta}\frac{e^{-i\gamma}+P/T}{e^{i\gamma}+P/T}\,.
\end{equation}
In the limit where the penguin-to-tree ratio $P/T$ is zero this reduces to
$S_{\pi\pi}=\sin2\alpha$. Although $P/T$ is highly suppressed, it is
not negligible and can be expressed, in QCD factorisation,
in terms of the CKM parameters and the
Gegenbauer moments $a_n$. Neglecting the small contributions from
weak annihilation terms \cite{QCDfac}, we find that $P/T$ is given 
by a ratio of polynomials
in $a_n$ with complex coefficients. For the asymptotic
DA, for instance, one has
\begin{equation}
  \frac{P}{T}=-\frac{1}{R_b}
\frac{(-1.36-0.31i)-(1.46+0.37i)\,a_0+(0.22+0.01i)\,a_0^2}{(13.0-0.08i)+
(15.5+0.03i)\,a_0-(0.56-0.15i)\,a_0^2}\,,
\end{equation}
where $R_b = \sqrt{\bar{\rho}^2+\bar{\eta}^2}= 0.40$ \cite{UTfit}.

The variation of $S_{\pi\pi}$ in terms of $a$ and $\Delta$ is shown in
Fig.~\ref{fig:BRpipi}.  For  $1\leq
\Delta\leq 1.2$ as suggested in Sec.~\ref{sec3}, the variation is of
order 1\% and becomes more significant only for unrealistically large
values of $\Delta$; the convergence for
$\Delta=1$ corresponds to the asymptotic wave function.  The current
experimental results \cite{CPpipi} are
\begin{equation}
  S_{\pi\pi} = -0.30\pm0.17\pm0.03 \hspace{10pt}(\textrm{BaBar}), \qquad   
S_{\pi\pi}=-1.00\pm0.21\pm0.07  \hspace{10pt}(\textrm{Belle}),
\end{equation}
which could only be accommodated using very extreme values of
$\Delta$. For example, taking a model with same-sign fall-off, to reproduce
the BaBar result with $a=2$ would require a value of $\Delta>10$ to be within
the $1\sigma$ band and $\Delta=20$ to approach the central value.
Even for higher values of $a$, a minimum $\Delta\approx 16$ is needed to 
approach $S_{\pi\pi}=-0.3$, and would
produce Gegenbauer moments that are significantly outside the known
constraints, for example with $a=6$ and $\Delta=16$, we find 
$a_2(2.2$ GeV$)=7.7$
and $a_4(2.2$ GeV$)=0.5$.  Obtaining a value of $S_{\pi\pi}=-1.0$ is not
possible 
for values of $\Delta\geq1$, which is required to keep $a_2$ positive. 

The effect of the DA model on the branching ratios of $B\to\pi^+\pi^-$ is
significantly more pronounced than for the CP-asymmetry.  The central value
of the branching ratio for the asymptotic DA is
\begin{equation}
{\rm BR}(B\to\pi^+\pi^-)=
5.5\times10^{-6}|0.25e^{i\cdot15^\circ} + e^{-i\gamma}|^2,
\end{equation}
where $\gamma= 60^\circ\pm7^\circ$ \cite{UTfit}, and the
explicit dependence of the branching ratio on the Gegenbauer moments is again a
polynomial in $a_n$:
\begin{eqnarray}
    {\rm BR}(B\to\pi^+\pi^-)&\approx&1.50\times10^{-6}\,|1+(1.17-0.01i)\,a_0 -
      (0.06+0.01i)\,a_0^2\,\\
      && +\,(1.16-0.01i)\,a_2-(0.02+0.01i)\,a_2^2-
(0.08+0.02i)\,a_0\,a_2\,+\dots|^2,\nonumber
\end{eqnarray}
where the terms of first order in $a_n$ come from nonspectator
interaction and those of second order from the hard-gluon exchange.
Figure {\ref{fig:BRpipi}} shows the variation of the branching ratio for the
two model DAs, which has to be compared to 
the current experimental world average ${\rm BR}(B\to\pi^+\pi^-)=
(4.6\pm0.4)\times 10^{-6}$ \cite{HFAG}.  We see that large
increases from the asymptotic value are possible by increasing $\Delta$,
with values within the physical range, with a considerable effect of around
30\% possible with the alternating-sign fall-off model. On the other
hand, the effect of a nonasymptotic DA is to {\em increase} the
branching ratio, and hence leads the result of QCD factorisation even
further away from the experimental result.

\FIGURE{
$$\epsfxsize=0.45\textwidth\epsffile{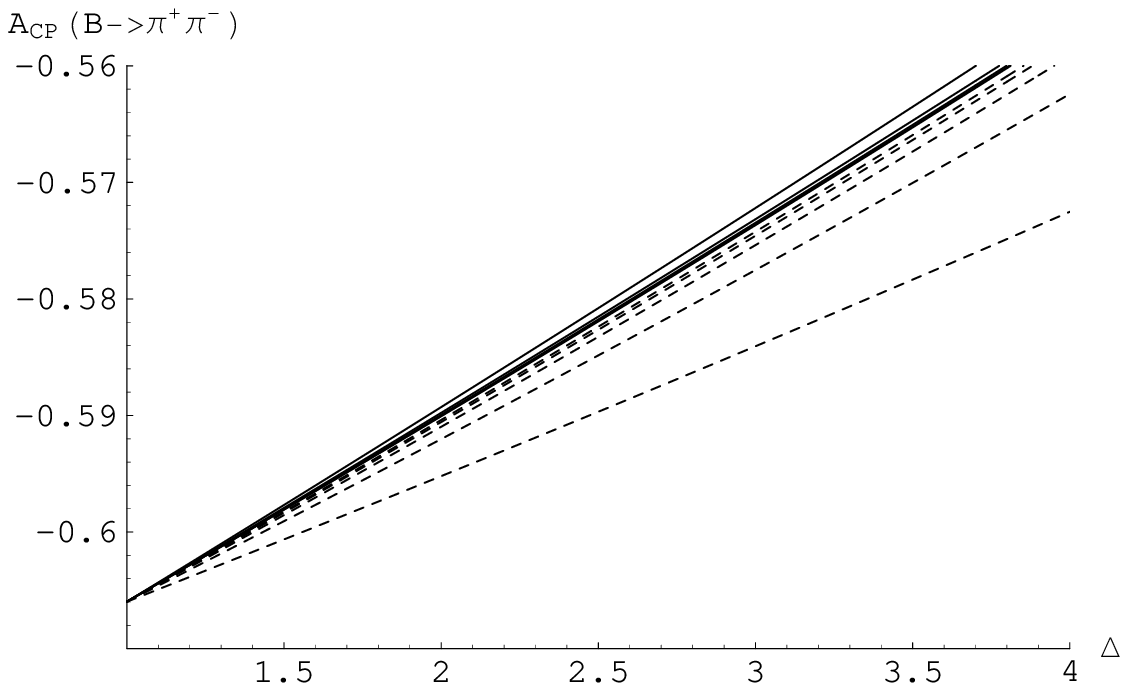}\qquad
\epsfxsize=0.45\textwidth\epsffile{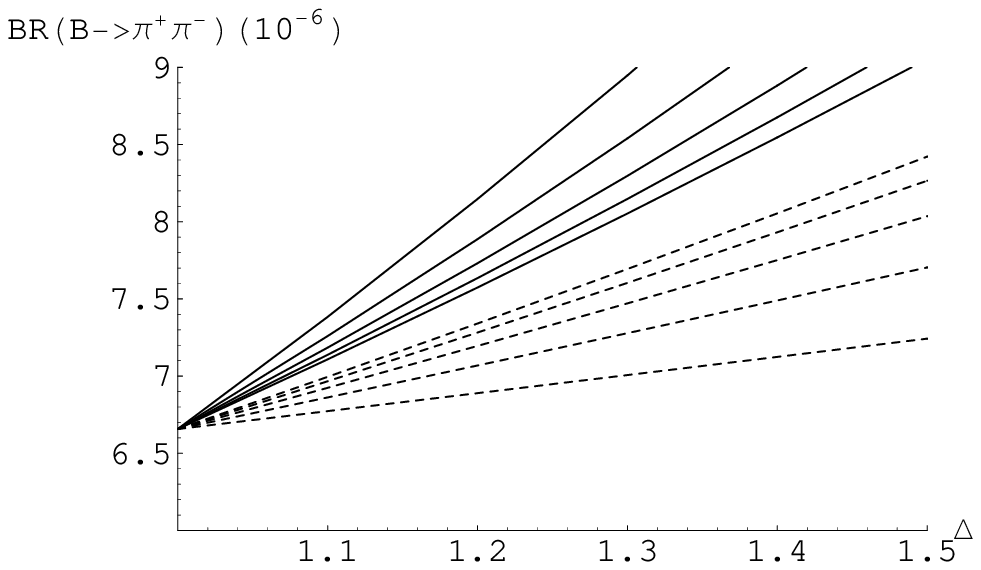}$$
\caption[]{Left: $A_{CP}(\pi\pi)$ as a function of $\Delta$
 for $a=2,3\dots6$, shown for models DAs with same-sign fall-off (dashed
 lines) and alternating sign fall-off (solid lines). Right: the same for 
${\rm BR}(B\to\pi^+\pi^-)$. Both sets of curves converge for
 increasing values of $a$.}
\label{fig:BRpipi}}

The situation for $B\to\pi K$ decays is complicated by the presence of the $K$
DA $\phi_K$, for which very little is known about the Gegenbauer
moments.  As a result of this, we use the same moments for $\phi_K$ as in
$\phi_\pi$ with the addition of the parameter $a_1^K$, for which we use the
value of $a_1^K(1.2$ GeV$)=0.15$, as discussed in Sec.~\ref{sec3}.  
Concentrating on the decay $\bar B_0\to\pi^+K^-$ and its CP-conjugate, 
we consider first the direct CP asymmetry, recently reported
in Ref.~\cite{CPKpi} as
\begin{eqnarray}
  A_{CP}^{\pi K}&=&
-0.133\pm0.030\pm0.009 \hspace{7pt}(\textrm{BaBar}), \nonumber\\
  A_{CP}^{\pi K}&=&-0.101\pm0.025\pm0.005
  \hspace{7pt}(\textrm{Belle}).
\label{exp}
\end{eqnarray}
In the framework of QCD factorisation, $A_{CP}$ can
be written in terms of real, strong interaction parameters (derived from
the factorisation coefficients) and pure CKM variables.  We can neglect the
annihilation contributions, but the electroweak penguins play a crucial role
and must therefore be included, which yields
\begin{eqnarray}
  \label{eq:AcppiK}
  A_{CP}^{\pi K} = \frac{\tan^2{\theta_c}\,R_b\,(\sin\gamma)\,r(\pi
  K)}{1+\tan^2{\theta_c}\,R_b\,(\cos\gamma)\,r^\prime(\pi K)},
\end{eqnarray}
where $\theta_c$ is the Cabibbo angle
and $r, r^\prime(\pi K)$ contain the QCD effects which depend on the
DA model parameters $\Delta$ and $a$.

Figure \ref{fig:ACPpiK} shows the dependence of the direct 
CP-asymmetry in $\bar B_0\to\pi^+ K^-$
on $\Delta$ and the fall-off parameter $a$. The first point to note is
that $A_{CP}$ differs in sign from the experimental value (\ref{exp}),
which is in agreement with the findings of \cite{QCDfac}. The precise
value of $A_{CP}$ does depend on $\Delta$, but it is impossible, even
for extreme DAs, to obtain the experimentally observed negative sign. 
This is also emphasised by
plotting the dependence of the asymmetry on the UT angle $\gamma$, as
expressed in (\ref{eq:AcppiK}), which is also shown 
Fig.~\ref{fig:ACPpiK}.  There is a 10\% increase in the value of
$A_{CP}(\pi K)$ between the asymptotic form with $\Delta=1$ 
and $\Delta=2$, more
significant changes only occurring outside the physical range of $\Delta$.
\FIGURE{
$$\epsfxsize=0.45\textwidth\epsffile{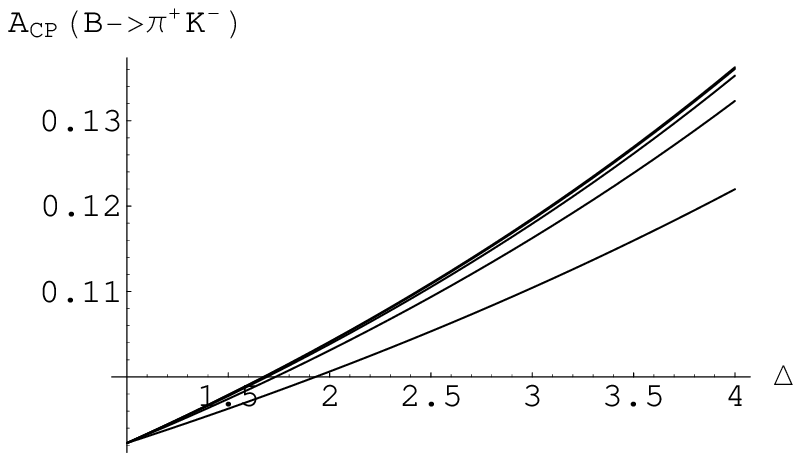}\qquad
\epsfxsize=0.45\textwidth\epsffile{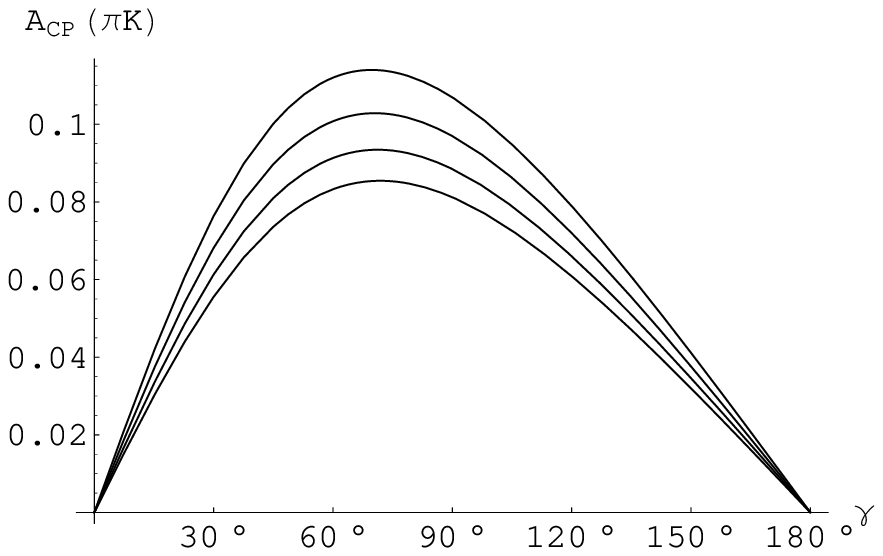}$$
\caption[]{Left: the variation of $A^{\pi K}_{CP}$ as a function of $\Delta$
        for $a=2,3\dots6$, for same-sign
       fall-off. Right: the dependence of $A^{\pi K}_{CP}$ on the UT 
angle $\gamma$ for curves at $a=2$, $\Delta
       = 1$ (lowest curve) to $\Delta = 4$.}
\label{fig:ACPpiK}}

As in the $B\to\pi^+\pi^-$ case, the effect of our model DAs on the branching
ratio of $\bar B_0\to\pi^+K^-$ 
is much more pronounced than for the CP-asymmetry and can exhibit up
to 30\% change from the asymptotic value within the physical ranges of $a$
and $\Delta$.  The branching ratio is given as
\begin{eqnarray}
  \textrm{BR}(B\to\pi K)\propto
  \left(F_0^{B\to\pi}(m_K^2)\right)^2 
|V_{cb}V_{cs}^*|^2\Big[\epsilon_{KM}e^{-i\gamma}c(\pi
  K) + c^\prime(\pi K)\Big]^2,
\end{eqnarray}
with $\epsilon_{KM}=\tan{\theta_c}^2R_b$ and $c, c^\prime(\pi K)$ are derived
from factorisation coefficients, containing all the dependence on
hadronic parameters and the DAs $\phi_{\pi,K}$.
Using the asymptotic DA we find $\textrm{BR}(B\to\pi
K)=13.58\times10^{-6}$, and the variation from this when higher-order
Gegenbauer moments are included is shown in Fig.~\ref{fig:BRpiK}.  
\FIGURE{
$$\epsfxsize=0.45\textwidth\epsffile{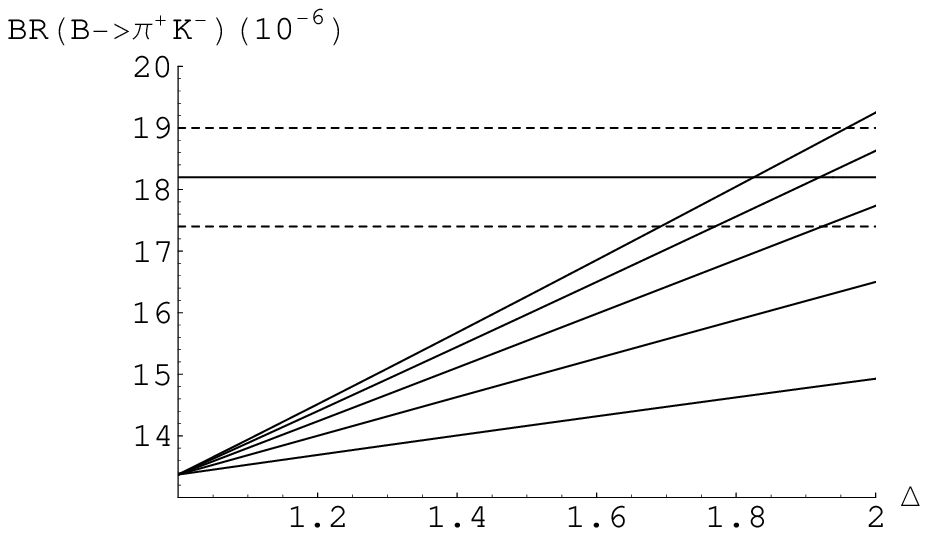}\qquad
\epsfxsize=0.45\textwidth\epsffile{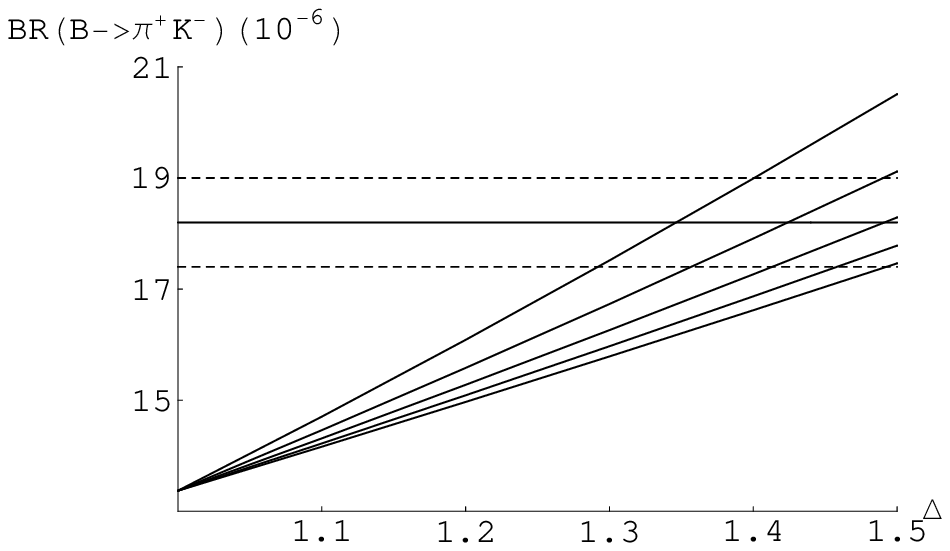}$$
\caption[]{$BR(B\to\pi^+K^-)$ as a 
function of $\Delta$
 for $a=2,3\dots6$, for models DAs with same-sign
 fall-off (left plot) and
 alternating sign fall-off (right plot).  
The experimental average is also marked for comparison.}
\label{fig:BRpiK}}
The experimental world average, also shown 
in Fig.~\ref{fig:BRpiK}, is $(18.2\pm0.8)\times 10^{-6}$ 
\cite{HFAG}. There
can be significant changes to the asymptotic value, especially in the model
with alternating sign fall-off.  The experimental average is shown
for comparison and can be
accommodated with $\Delta\approx1.8$ for the same sign fall-off, or
$\Delta\approx1.4$ with alternating sign fall-off. 
\FIGURE{
$$\epsfysize=0.32\textwidth\epsffile{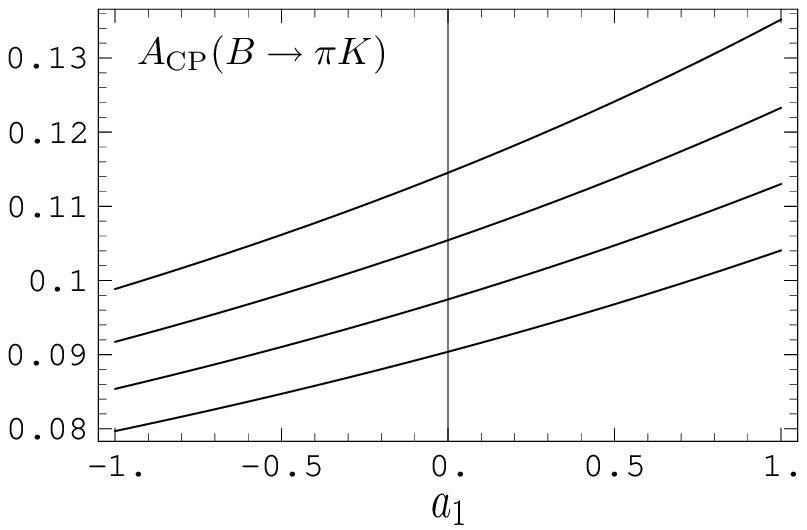}$$
\caption[]{The direct CP-asymmetry in $\bar B_0\to\pi^+ K^-$ as
  function of $a_1$ for $a=3$ and $\Delta\in\{1,2,3,4\}$ (from bottom
  to top).
$\gamma =  60^\circ$.}\label{fig:a1cp}
}
In
Fig.~\ref{fig:a1cp} we also show the dependence
of the direct CP\--asym\-me\-try in $B\to\pi K$ on 
$a_1$, the Gegenbauer moment that parameterises the antisymmetric part
of the $K$ DA, whose actual value is around $0.15$. It is
evident that one can shift $A_{CP}$ into the ``right'' direction by
decreasing $a_1$, but again extreme values of $a_1$ would be needed to
obtain a negative CP-asymmetry.
 
Our conclusion from this investigation is that the discrepancy between
the experimental values of BRs and CP\--asym\-me\-tries and their predicted
values in QCD factorisation can not be attributed to the
uncertainties in the leading-twist DAs: for $B\to\pi^+\pi^-$ it is
impossible to reproduce the data, whereas for the direct CP-\-asym\-me\-try in 
$B\to\pi K$ 
high\-ly unrealistic values of
$\Delta$ and $a$ would be needed in order to reconcile the theory
predictions with experimental data. 

\section{Summary and Conclusions}\label{sec6}

We have presented models for the symmetric part of 
leading-twist light-cone distribution
amplitudes (DAs) of light mesons which depend on three parameters. Two of
these parameters control the fall-off 
behaviour of the Gegenbauer moments $a_n$ in $n$,
whereas the third one, $\Delta$, is given by
the first inverse moment of the DA and
parameterises the maximum possible impact of higher Gegenbauer moments
on the actual physical amplitude described in factorisation. 
We have also developed
similar models for the antisymmetric part of
the DA, which is relevant for $K$ and $K^*$
mesons; these models are normalised to $a_1$, the first
Gegenbauer moment. For the $\pi$ DA,
for which experimental data exist, we have formulated constraints on
the model parameters which are likely to be valid also for other meson
DAs. We have argued
that these models are better suited to estimate the true hadronic
uncertainty of processes calculated in factorisation 
than the standard truncated
conformal expansion and have studied these uncertainties for two
quantities, the $B\to\pi$ weak decay form factor $f_+^{B\to\pi}(0)$
and the CP asymmetry in $B\to\pi\pi$ and $B\to K\pi$. For the
former, the theoretical uncertainty induced by the model-dependence of
the DA is smaller than that due to other parameters and
approximations. For nonleptonic decays calculated in QCD factorisation
we find that the branching ratios are more sensitive to the precise
values of the model parameters than the CP-asymmetries. In both decays
channels it is however impossible to explain the experimental data by
nonstandard DAs, which indicates the presence of nonnegligible
nonfactorisable contributions --- a conclusion that agrees with the
findings of other authors, e.g.\ Refs.~\cite{Feldmann,Buras}.

Our models should prove particularly useful for describing DAs of
mesons other than the $\pi$ which are also symmetric by virtue of
G-parity, but for which no experimental or other reliable theoretical
information is available --- in particular the $\rho$, $\omega$ and $\phi$. For
these particles, we argue that existing theoretical indications
from local QCD sum rules \cite{CZreport,BB96,SU(3)breaking} 
point at the DAs being
narrower than that of the $\pi$, which implies the allowed
values of $\Delta$ being smaller than 1.2 at the scale $1.2\,$GeV. 
The same results also imply $a_2$ 
to be positive, which entails $\Delta\geq 1$. For the parameter $a$,
which controls the fall-off of the $a_n$ in $n$, we have found that
the perturbative contributions to
QCD sum rules indicate it to be 3 \cite{wavefunctionsV}, but that also
smaller values of $a$ are not excluded unless one can rigorously prove
that the leading-twist DAs must behave as $\sim u(1-u)$ near the
endpoints $u=0,1$, at {\em all scales}, which would imply $a\geq 4$. 
One more relevant constraint for the models $\phi^-_{a,b}$ 
with an alternating-sign fall-off of the Gegenbauer moments comes from
the requirement $\phi_\pi(1/2)>0.9$ at the scale $\mu=1\,{\rm
  GeV}$. A positive value of this quantity is also required by 
QCD sum rules on the light-cone for the couplings
$g_{DD^*\pi}$ \cite{couplings1} and $g_{DD^*\rho}$ \cite{couplings2},
and hence also very likely to be the case for other mesons. 
But even without these
constraints being taken literally, our models provide a way to test
the impact of nonasymptotic DAs on physical amplitudes without
recourse to conformal expansion.

\acknowledgments

A.N.T.\ gratefully acknowledges receipt of a UK PPARC studentship.


\begin{thebibliography}{99} 

\bibitem{BZ2} 
P.~Ball and R.~Zwicky,
Phys.\ Rev.\ D {\bf 71} (2005) 014029
[arXiv:hep-ph/0412079].

\bibitem{pQCD}
V.~L.~Chernyak and A.~R.~Zhitnitsky,
JETP Lett.\  {\bf 25} (1977) 510
[Pisma Zh.\ Eksp.\ Teor.\ Fiz.\  {\bf 25} (1977) 544];\\
Sov.\ J.\ Nucl.\ Phys.\  {\bf 31} (1980) 544
[Yad.\ Fiz.\  {\bf 31} (1980) 1053];\\
A.V.\ Efremov and A.V.\ Radyushkin,
Phys.\ Lett.\ B {\bf 94} (1980) 245;
Theor.\ Math.\ Phys.\  {\bf 42} (1980) 97
[Teor.\ Mat.\ Fiz.\  {\bf 42} (1980) 147];\\
G.P.\ Lepage and S.J.\ Brodsky,
Phys.\ Lett.\ B {\bf 87} (1979) 359;
Phys.\ Rev.\ D {\bf 22} (1980) 2157;\\
V.L.\ Chernyak, A.R.\ Zhitnitsky and V.G.\ Serbo,
JETP Lett.\  {\bf 26} (1977) 594
[Pisma Zh.\ Eksp.\ Teor.\ Fiz.\  {\bf 26} (1977) 760];
Sov.\ J.\ Nucl.\ Phys.\  {\bf 31} (1980) 552
[Yad.\ Fiz.\  {\bf 31} (1980) 1069].

\bibitem{BLreport} S.J.\ Brodsky and G.P.\ Lepage, in: {\em Perturbative
    Quantum Chromodynamics}, ed.\ by A.H.~Mueller, p.~93, World
  Scientific (Singapore) 1989.
                                                                       
\bibitem{LCSRs:reviews}
P. Colangelo and A. Khodjamirian,
hep-ph/0010175;\\
A. Khodjamirian,
hep-ph/0108205.

\bibitem{protz}
V.M.\ Belyaev, A. Khodjamirian and R. R\"uckl,
Z.\ Phys.\ C {\bf 60} (1993) 349
[hep-ph/9305348];\\
P.~Ball and V.~M.~Braun,
Phys.\ Rev.\ D {\bf 55} (1997) 5561
[arXiv:hep-ph/9701238];\\
A. Khodjamirian {\it et al.},
Phys.\ Lett.\ B {\bf 410} (1997) 275
[hep-ph/9706303];\\
E. Bagan, P. Ball and V.M.\ Braun,
Phys.\ Lett.\ B {\bf 417} (1998) 154
[hep-ph/9709243];\\
P. Ball,
JHEP {\bf 9809} (1998) 005
[hep-ph/9802394];\\
P.~Ball and V.~M.~Braun,
Phys.\ Rev.\ D {\bf 58} (1998) 094016
[arXiv:hep-ph/9805422];\\
A. Khodjamirian {\it et al.},
Phys.\ Rev.\ D {\bf 62} 114002 (2000)
[hep-ph/0001297].

\bibitem{BZ}
P.~Ball and R.~Zwicky,
JHEP {\bf 0110} (2001) 019
[arXiv:hep-ph/0110115];\\
P.~Ball and R.~Zwicky,
Phys.\ Rev.\ D {\bf 71}, 014015 (2005)
[arXiv:hep-ph/0406232].

\bibitem{QCDfac}
M.~Beneke {\em et al.},
Phys.\ Rev.\ Lett.\  {\bf 83} (1999) 1914
[arXiv:hep-ph/9905312];\\
M.~Beneke {\em et al.},
Nucl.\ Phys.\ B {\bf 606} (2001) 245
[arXiv:hep-ph/0104110];\\
M.~Beneke and M.~Neubert,
Nucl.\ Phys.\ B {\bf 675} (2003) 333
[arXiv:hep-ph/0308039].

\bibitem{SCET}
C.~W.~Bauer {\em et al.},
Phys.\ Rev.\ D {\bf 63} (2001) 114020
[arXiv:hep-ph/0011336].

\bibitem{CZreport}
V.~L.~Chernyak and A.~R.~Zhitnitsky,
Phys.\ Rept.\  {\bf 112} (1984) 173.

\bibitem{wavefunctionsPS}
V.~M.~Braun and I.~E.~Filyanov,
Z.\ Phys.\ C {\bf 48} (1990) 239
[Sov.\ J.\ Nucl.\ Phys.\  {\bf 52} (1990) 126];\\
P. Ball,
JHEP {\bf 9901} (1999) 010
[hep-ph/9812375].

\bibitem{wavefunctionsV}
P. Ball {\em et al.}, 
Nucl.\ Phys.\ B {\bf 529} (1998) 323
[arXiv:hep-ph/9802299];\\
P.~Ball and V.~M.~Braun,
Nucl.\ Phys.\ B {\bf 543} (1999) 201
[arXiv:hep-ph/9810475].

\bibitem{CLEO}
J.~Gronberg {\it et al.}  [CLEO Collaboration],
Phys.\ Rev.\ D {\bf 57} (1998) 33
[arXiv:hep-ex/9707031].
    \bibitem{lattDA}
G.~Martinelli and C.~T.~Sachrajda,
Phys.\ Lett.\ B {\bf 190} (1987) 151;\\
T.~A.~DeGrand and R.~D.~Loft,
Phys.\ Rev.\ D {\bf 38} (1988) 954;\\
D.~Daniel, R.~Gupta and D.~G.~Richards,
Phys.\ Rev.\ D {\bf 43} (1991) 3715.

\bibitem{sachrajda}
L.~Del Debbio {\em et al.} [UKQCD Collaboration],
Nucl.\ Phys.\ Proc.\ Suppl.\  {\bf 83} (2000) 235
[arXiv:hep-lat/9909147];\\
L.~Del Debbio, M.~Di Pierro and A.~Dougall,
Nucl.\ Phys.\ Proc.\ Suppl.\  {\bf 119} (2003) 416
[arXiv:hep-lat/0211037].
         
\bibitem{Filyanov}
V.~M.~Braun and I.~E.~Filyanov,
Z.\ Phys.\ C {\bf 44}, 157 (1989)
[Sov.\ J.\ Nucl.\ Phys.\  {\bf 50}, 511 (1989)].

\bibitem{SU(3)breaking}
P.~Ball and M.~Boglione,
Phys.\ Rev.\ D {\bf 68} (2003) 094006
[arXiv:hep-ph/0307337].

\bibitem{BB96}
P.~Ball and V.~M.~Braun,
Phys.\ Rev.\ D {\bf 54} (1996) 2182
[arXiv:hep-ph/9602323].

\bibitem{nonlocal}
S.~V.~Mikhailov and A.~V.~Radyushkin,
Phys.\ Rev.\ D {\bf 45} (1992) 1754.

\bibitem{Stefanis2}
A.~P.~Bakulev, S.~V.~Mikhailov and N.~G.~Stefanis,
Phys.\ Lett.\ B {\bf 508} (2001) 279
[Erratum-ibid.\ B {\bf 590} (2004) 309]
[arXiv:hep-ph/0103119].

\bibitem{bigi}
I.~I.~Bigi,
arXiv:hep-ph/0501084.

\bibitem{Schmedding}
A.~Schmedding and O.~I.~Yakovlev,
Phys.\ Rev.\ D {\bf 62} (2000) 116002
[arXiv:hep-ph/9905392].

\bibitem{Stefanis1}
A.~P.~Bakulev {\em et al.},
Phys.\ Rev.\ D {\bf 70} (2004) 033014
[Erratum-ibid.\ D {\bf 70} (2004) 079906]
[arXiv:hep-ph/0405062].

\bibitem{GW}
D.~J.~Gross and F.~Wilczek,
Phys.\ Rev.\ D {\bf 9} (1974) 980;\\
M.~A.~Shifman and M.~I.~Vysotsky,
Nucl.\ Phys.\ B {\bf 186}, 475 (1981).

\bibitem{first}
V.~L.~Chernyak and I.~R.~Zhitnitsky,
Nucl.\ Phys.\ B {\bf 345} (1990) 137.

\bibitem{newphys}
For instance:\\
V.~Barger {\it et al.},
Phys.\ Lett.\ B {\bf 598} (2004) 218
[arXiv:hep-ph/0406126];\\
S.~Khalil and E.~Kou,
arXiv:hep-ph/0407284;\\
S.~Mishima and T.~Yoshikawa,
Phys.\ Rev.\ D {\bf 70} (2004) 094024
[arXiv:hep-ph/0408090];\\
A.~J.~Buras {\em et al.},
arXiv:hep-ph/0410407;\\
S.~Baek {\em et al.},
arXiv:hep-ph/0412086.

\bibitem{Feldmann}
T.~Feldmann and T.~Hurth,
JHEP {\bf 0411} (2004) 037
[arXiv:hep-ph/0408188].

\bibitem{UTfit}
M.~Bona {\it et al.}  [UTfit Collaboration],
arXiv:hep-ph/0501199.

\bibitem{CPpipi}
B.~Aubert {\it et al.}  [BABAR Collaboration],
[arXiv:hep-ex/0408089];\\
K.~Abe {\it et al.}  [Belle Collaboration],
Phys.\ Rev.\ Lett.\  {\bf 93} (2004) 021601
[arXiv:hep-ex/0401029].

\bibitem{HFAG}
H.~F.~A.~Group,
arXiv:hep-ex/0412073.

\bibitem{CPKpi}
B.~Aubert {\it et al.}  [BaBar Collaboration],
Phys.\ Rev.\ Lett.\  {\bf 93} (2004) 131801
[arXiv:hep-ex/0407057];\\
Y.~Chao {\it et al.}  [Belle Collaboration],
Phys.\ Rev.\ Lett.\  {\bf 93} (2004) 191802
[arXiv:hep-ex/0408100].

\bibitem{couplings1}
V.~M.~Belyaev {\em et al.},
Phys.\ Rev.\ D {\bf 51} (1995) 6177
[arXiv:hep-ph/9410280];\\
A.~Khodjamirian {\em et al.},
Phys.\ Lett.\ B {\bf 457} (1999) 245
[arXiv:hep-ph/9903421].

\bibitem{couplings2}
Z.~H.~Li {\em et al.},
Phys.\ Rev.\ D {\bf 65} (2002) 076005
[arXiv:hep-ph/0208168].

\bibitem{Buras}
A.~J.~Buras {\em et al.},
Nucl.\ Phys.\ B {\bf 697} (2004) 133
[arXiv:hep-ph/0402112].

\end{thebibliography}
\end{document}